\documentclass[pre,floats,superscriptaddress,usenames]{revtex4}

\usepackage{amsmath}
\usepackage{amssymb}
\usepackage{epsfig}
\usepackage{multirow}
\usepackage{graphicx}
\newcommand{\be}{\begin{equation}} 
\newcommand{\ee}{\end{equation}}
\newcommand{\bea}{\begin{eqnarray}}   
\newcommand{\eea}{\end{eqnarray}}

\newcommand{\rr}{{\bf r}}
\newcommand{\NN}{{\bf \nabla}}
\newcommand{\FF}{{\bf F}}
\newcommand{\GG}{{\bf G}}

\newcommand{\vv}{{\bf v}}
\newcommand{\vva}{{\bf v}^{\alpha}}
\newcommand{\vvb}{{\bf v}^{\beta}}
\newcommand{\vvab}{{\bf v}_{\alpha\beta}}
\newcommand{\fa}{f^{\alpha}}
\newcommand{\na}{n^{\alpha}}
\newcommand{\nb}{n^{\beta}}
\newcommand{\rhoa}{\rho^{\alpha}}

\newcommand{\fb}{f^{\beta}}

\newcommand{\uu}{{\bf u}}
\newcommand{\uua}{{\bf u}^{\alpha}}
\newcommand{\ua}{u^{\alpha}}

\newcommand{\ub}{u^{\beta}}
\newcommand{\uub}{{\bf u}^{\beta}}
\newcommand{\ma} {m^{\alpha}}

\newcommand{\vaj} {v^{\alpha }_j}
\newcommand{\vai} {v^{\alpha }_i}

\newcommand{\mb} {m^{\beta}}
\newcommand{\sab}{\sigma_{\alpha\beta}}
\newcommand{\gab}{g_{\alpha\beta}}
\newcommand{\muab}{\mu_{\alpha\beta}}
\newcommand{\bk}{\hat{\bf s}}

\begin{document}
\date{\today}

%\markboth{Umberto Marini Bettolo Marconi}
%\articletype{RESEARCH ARTICLE}

\title{Non-local kinetic theory of inhomogeneous liquid mixtures}
\author{Umberto Marini Bettolo Marconi\\
{(umberto.marinibettolo@unicam.it)}
}

\address{ Scuola di Scienze e Tecnologie, 
Universit\`a di Camerino, Via Madonna delle Carceri, 62032 ,
Camerino, INFN Perugia, Italy}
\begin{abstract}
In this work we investigate
the dynamical properties of  a mixture
of mutually interacting spherical molecules of different masses and sizes. 
From an analysis of the  microscopic laws governing the motion of the
molecules we derive a set of non-local self-consistent  equations for the singlet phase-space distribution functions.
The theory  is shown to reproduce the
hydrodynamic equations for the densities of each species, the total momentum and the local temperature.
 The non ideal gas interaction term is separated into a contribution due to the repulsive part, which is treated by
means of the revised Enskog theory for hard spheres, and an attractive contribution
treated within the random phase approximation. The present formulation accounts for
the effects of the density and velocity inhomogeneities both on the thermodynamic and transport 
properties of the fluid. 

In a special limit, where one species is 
massive and diluted, the theory leads to a description 
 which is formally identical to the dynamic density
functional equation governing the time evolution of  a colloidal system.
 The derivation also determines the dependence  of  the friction coefficient, appearing
 in the dynamic density functional theory,
  on the microscopic parameters of the solvent. However,  the predicted value
takes into account only the collisional contributions to the friction
and not the Stokes friction of hydrodynamic origin, suggesting that 
velocity correlations should be incorporated in a more complete treatment.
 %\begin{keywords}
%Kinetic theory, multi-component diffusion, dynamical density functional theory
%\end{keywords}\bigskip
\end{abstract}
\maketitle
\section{Introduction}

  Spatially inhomogeneous systems on mesoscopic length scales can generate properties 
which do not appear in bulk materials, offering new perspectives for future applications.
A variety of tools have been utilized  to study the properties
of fluids near substrates and  liquid interfaces  in terms of molecular forces, 
ranging from new experimental techniques to
numerical algorithms and theoretical approaches \cite{Squires,schoch,sparreboom,Evans2,rauscher,Bruus}.

It is an  understatement to say that, since its appearance in 1979, the {\it Manifesto} \cite{Evans1} of density functional
theory (DFT)    has  strongly influenced the studies in the field  of classical inhomogeneous fluids.
In this approach  the equilibrium density profile is determined by a functional derivative of a non-local
Helmholtz free energy functional. An exact theorem states that such an equilibrium
profile minimizes the grand potential of the system and is unique. This fact renders the method extremely
appealing and provides a great help in finding good approximate Helmholtz functionals in a variety of cases.

Regarding out of equilibrium systems, we do not have such useful theorems  \cite{Fink} and 
a time dependent extension of these methods, the dynamical  density functional theory (DDFT), has been applied on 
phenomenological grounds \cite{ArcherEvans,Tarazona}. As shown by means of comparisons with  numerical simulations, 
DDFT is able to capture
the over-damped dynamics of suspensions, but does not account for
the richer dynamics of molecular fluids \cite{Archer2005,Archer2009,rauscher}. Although some hydrodynamic aspects, such as 
the presence of a drift in the solvent or  inertial corrections  \cite{pedro,Cecconi,Melchionna2007}, can
be included in the DDFT, a full treatment of these effects requires a different approach.
In order to capture the
isothermal hydrodynamic behavior,  describing the long length and long time scale behavior of fluids, one must consider
 the momentum  density in addition to  the mass density. 
Hydrodynamics, developed much earlier than Kinetic theory and without any knowledge of the underlying
microscopic structure of liquids, but using only phenomenological arguments and  conservation laws \cite{Hydrodynamics},
represents a universal theory of fluid behavior.
With this remark we want to stress the fact
 that, in constructing a dynamical description of a fluid valid both at molecular and  macroscopic scales,
we must be consistent  with the assumptions which rendered classical hydrodynamics such a
successful theory of liquids. In other words, symmetries and mass and  momentum conservation laws
must be exactly preserved even when necessary and unavoidable approximations are introduced.

When a liquid is treated  as a mechanical continuum, it is relatively simple to find 
a closed set of governing hydrodynamic equations for the relevant fields
with the help of the so-called phenomenological constitutive relations.
However, when  the physical inhomogeneities become of the same order of magnitude as the molecular scales,
one is forced to adopt a microscopic description \cite{Huang}. 
The challenge is to extend macroscopic concepts  such as pressure, viscosity, diffusivity and thermal conductivity
to situations, where matter is confined to narrow spaces or is heterogeneous.
All these properties assume a non-local dependence
on the controlling  fields and this fact renders the theory particularly challenging.

The present paper is organized as follows: in Sec.
\ref{Theory} we introduce the model and the evolution
equations for the distribution functions of the
multi-component system. 
Numerical solutions of the transport equations are possible
by means of the Lattice Boltzmann technique \cite{LBgeneral} as discussed elsewhere  \cite{Melchionna2008,Melchionna2009,Lausanne2009,JCP2010}.
In Sec. \ref{Equilibrium properties} 
 we analyze the predictions of the theory concerning the equilibrium and non equilibrium
 behavior of the model and give the relevant formulae to compute the pressure and the
 transport coefficients.
  In Sec. \ref{DDFT} we mimic a colloidal suspension by  specializing the discussion to a 
 hard-sphere binary mixture, composed
 of a low-concentration heavy species and a high concentration 
 light species.
 From this non primitive model we 
 obtain an equation for the evolution of the heavy species alone
 very similar to the DDFT equation. This heuristic derivation also
shows the limits of approaches based on 
 the neglect of velocity correlations.
Finally in Sec. \ref{Conclusions} we present some conclusions
and perspectives.

%%%%%%%%%%%%%%%%%%
\section{Theory}
\label{Theory}
We consider  an M-component fluid, whose species, denoted  by 
the label $\alpha=1,M$, have masses $m_\alpha$ and interact with pair additive, centrally symmetric potentials $U^{\alpha\beta}(r)$
and  external forces $\FF_\alpha$.
We  describe the statistical evolution  of the system by means of the so-called BBGKY hierarchy of equations \cite{hansen},
whose first  $M$ equations involve the  singlet
phase space distribution functions, $\fa(\rr,\vv,t)$ and the two particle distribution functions,
 $f_2^{\alpha\beta}(\rr,\rr',\vv,\vv',t) $ :

\bea
\partial_{t}\fa(\rr,\vv,t) +\vv\cdot\NN \fa(\rr,\vv,t)
+\frac{\FF^{\alpha}(\rr)}{\ma}\cdot
\frac{\partial}{\partial \vv} \fa(\rr,\vv,t)= \sum_\beta\Omega^{\alpha\beta}(\rr,\vv,t) 
\label{evolution}
\eea
where the interaction term $\Omega^{\alpha\beta}$ is given by:
\be
\Omega^{\alpha\beta}(\rr,\vv,t)=
\frac{1}{\ma}\NN_v \cdot \int d\vv' \int dr'
\NN_r U^{\alpha\beta}(\rr-\rr')f_2^{\alpha\beta}(\rr,\rr',\vv,\vv',t) .  
\label{coll}
\ee

The
evolution equations for the singlet distributions  involve the
two particle distributions and  these in turn involve the three-particle distributions.   
In order to  make some progress one needs 
to approximate these higher order distributions, using some physically motivated prescription.
A popular closure ansatz allowing to reduce the BBGKY equations \cite{hansen} to a closed set 
of equations for the singlet distributions is the following:
\be
f_2^{\alpha\beta}(\rr,\rr',\vv,\vv',t)\simeq\gab(\rr,\rr',t)f^{\alpha}(\rr,\vv,t)f^{\beta}(\rr',\vv',t),
\label{factorization}
\ee
where $\gab(\rr,\rr',t)$ is the  local  equilibrium pair distribution function. 
The approximation (\ref{factorization}) incorporates static correlations correctly, but neglects velocity correlations,
through the lack of velocity dependence in $\gab(\rr,\rr',t)$. We shall comment later on
the consequences of such an ansatz. We assume that the interaction potential between two particles can be separated into a
short-range strongly repulsive and a longer range attractive contribution, so that the interaction 
process results in a combination  of almost instantaneous,  hard core collisions and small velocity 
changes induced by the weak attractive field:
\be
\Omega^{\alpha\beta}(\rr,\vv,t)=\Omega^{\alpha\beta}_{rep}(\rr,\vv,t)+
\Omega^{\alpha\beta}_{att}(\rr,\vv,t).
\label{opcoll}
\ee
The repulsive part of $\Omega_{rep}^{\alpha\beta}$ is treated using the Revised Enskog theory of Ernst and van Beijeren
\cite{VanBeijeren,Lopezdeharo,Stell,Kincaid,Stell2}
for hard-sphere mixtures of diameters $\sigma_{\alpha\beta}$,
which  neglects velocity correlations for two particles about to collide, as in Boltzmann theory, but
includes the configurational correlations resulting from the finite size of the particles via the pair correlation function at contact:
\bea
&&\Omega_{rep}^{\alpha\beta}(\rr,\vva,\vvb,t)
= \sab^2\int d\vvb\int 
d\bk\Theta(\bk\cdot \vvab) (\bk 
\cdot \vvab)\times\nonumber\\
&&\Bigl\{ \gab
(\rr,\rr-\sab\bk,t) \fa (\rr,{\bar\vv}^\alpha,t)\fb (\rr-\sab\bk,{\bar\vv}^\beta,t)
\nonumber\\&& 
 -\gab(\rr,\rr+\sab\bk,t)\fa(\rr,\vva,t)\fb(\rr+\sab\bk,\vvb,t)\Bigl\},
\label{ret}
\eea
where $\Theta(x)$ is the Heaviside function,  $\vv_{\alpha\beta}=(\vva-\vvb)$, while ${\bar\vv}^\alpha$ and ${\bar\vv}^\beta$ are scattered velocities given by
\bea
&&
{\bar\vv}^\alpha=\vva-\frac{2 \mb}{\ma+\mb}(\bk\cdot\vv_{\alpha\beta})\bk\nonumber\\
&&
{\bar\vv}^\beta=\vvb+
\frac{2 \ma}{\ma+\mb}(\bk\cdot\vv_{\alpha\beta})\bk
\label{collrule}
\eea
and 
$\bk$ is the unit vector directed from 
particle $\alpha$ to particle $\beta$. 
The quantities $\gab(\rr,\rr\pm\sab\bk)$ are the inhomogeneous
hard sphere pair correlation functions evaluated when the particles
of species $\alpha$ and $\beta$  are at contact distance $\sigma_{\alpha\beta}= (\sigma_{\alpha\alpha}+\sigma_{\beta\beta})/2$. 
The  attractive term has the random phase approximation \cite{hansen} expression:
\bea
&&
 \Omega^{\alpha\beta}_{attr}(\rr,\vv,t)=
-\frac{\GG^{\alpha\beta}(\rr,t)}{\ma}\cdot \NN_v \fa(\rr,\vv,t)
\label{rpa}
\eea
where $\GG^{\alpha\beta}$ are  the molecular fields
\be
\GG^{\alpha\beta}(\rr,t)=
- \int d\rr' \nb(\rr',t)\gab(\rr,\rr',t)\NN_r U_{attr}^{\alpha\beta}(\rr-\rr')
\ee
and $\nb$ the number density of species $\beta$.
Even with  approximations \eqref{ret} and \eqref{rpa} the full solution of eq. \eqref{evolution}
is exceedingly difficult and can be found only in some special cases of limited interest.
In order to encompass this problem
many authors adopted  simpler forms of the  interaction term. 
Among these forms,  a very popular recipe is represented by the BGK model \cite{BGK} consisting in replacing $\Omega^{\alpha\beta}$ by
a relaxation term. For collisions between particles belonging to the same species one chooses:
\be
\Omega^{\alpha\alpha}_{BGK}=-\omega_{\alpha\alpha}(\fa(\rr,\vv,t)-\psi^{\alpha}(\rr,\vv,t))
\ee
and for collisions between unlike particles:
\be
\Omega^{\alpha\bar\alpha}_{BGK}=-\omega_{\alpha\bar\alpha}(\fa(\rr,\vv,t)-\bar\psi^{\alpha}(\rr,\vv,t))
 \ee
 where $\psi^{\alpha}$ is the local equilibrium distribution for species $\alpha$:
 \be
\psi^{\alpha}(\rr,\vv,t)=\na(\rr,t)[\frac{\ma}{2\pi k_B T}]^{3/2}\exp
\Bigl(-\frac{\ma(\vv-\uu(\rr,t))^2}{2 k_B T} \Bigl).
\ee 
and the functions $\bar\psi^{\alpha}$ and the frequencies $\omega_{\alpha\bar\alpha}$ 
have to be modeled to account for  collisions between particles of different species \cite{Sofonea}.

The  BGK approximation is based on the idea that the system after few molecular collisions reaches
a state of local thermodynamic equilibrium, where
the distributions depend on space and time only through the hydrodynamic variables of the system,
$\rhoa(\rr,t),\uu(\rr,t),T(\rr,t)$, partial mass densities, average velocity and temperature, respectively.
These variables are  defined in terms of the distributions as:
\be
\rhoa(\rr,t)=\ma \na(\rr,t) = \ma \int d\vv \fa(\rr,\vv,t),
\label{density}
\ee
\be
\uua(\rr,t)=\frac{1}{\na(\rr,t)} \int d\vv \vv \fa(\rr,\vv,t),
\label{momentum}
\ee
and
\be
T(\rr,t)=
\frac{1}{3 n (\rr,t)}
\sum_\alpha \ma \int d\vv (\vv-\uu)^2 \fa(\rr,\vv,t) 
\ee
with total density given by $n(\rr,t)=\sum_\alpha \na(\rr,t)$,
barycentric velocity by $\uu(\rr,t)=\sum_\alpha \rhoa(\rr,t)\uua(\rr,t)/\rho(\rr,t)$ and global density $\rho(\rr,t)=\sum_\alpha\rhoa(\rr,t)$.
With the help of the  BGK ansatz one easily derives the hydrodynamic equations for the conserved variables
$\rhoa(\rr,t),\rho \uu(\rr,t)$ and the energy
and via the Chapman-Enskog analysis \cite{Chapman} the transport coefficients.
Unfortunately, the method gives an oversimplified picture of the thermodynamic properties
of the system, which turn out to be  the same as those of an ideal gas, since 
$\Omega^{\alpha\beta}_{BGK}$
does not contribute to the pressure or to the surface tension. A partial remedy to such a situation was introduced by
Shan and Chen and other authors \cite{shanchen,He,Doolen}. To take into account the contribution of the interactions
to the equation of state they included a self-consistent term,  named the pseudo-potential. 
This ad-hoc adjustment allowed to consider the hydrodynamic properties of non ideal gases
by means of the so-called Lattice Boltzmann method (LBM) \cite{LBgeneral}.
The pseudo-potential is a 
conservative force and does not determine the  transport coefficients, which only depend on the 
adjustable parameters $\omega_{\alpha\beta}$. In the language of the present article the Shan-Chen method
is a particular choice of $\GG^{\alpha\beta}$ with the peculiarity that it contains both attractive
and repulsive contributions, an assumption which is somehow in conflict  with the standard methods
of liquid state theory, where one treats separately and on a different basis attractive and repulsive forces \cite{WCA}.
 
The full collision RET operator \eqref{ret} being a non-linear functional of $\fa$,
couples the different velocity moments of the distributions thus rendering analytic work very hard,
unless one adopts a suitable truncation scheme.

Any satisfactory approximation
must retain the physical symmetries and   conservation laws which are incorporated in the microscopic representation
\eqref{ret} and \eqref{rpa}.
To achieve that goal Dufty et al. \cite{Brey,Brey1}
separated the contributions of $\Omega^{\alpha\beta}$ to the hydrodynamic equations from
those affecting the evolution of non-hydrodynamic modes,
by projecting the collision term onto the hydrodynamic subspace spanned by the functions $\{1,\vv,v^2\}$
and onto the complementary kinetic subspace:
 
\be
\Omega^{\alpha\beta}={\cal P}_{hydro}\Omega^{\alpha\beta} +(I-{\cal P}_{hydro})\Omega^{\alpha\beta}
\label{splitting}
\ee
with 
%%%%%%%%%%%%%%%%%%%%%%%
\be
{\cal P}_{hydro}   \Omega^{\alpha\beta}(\rr,\vv,t)\equiv
\frac{1}{k_B T(\rr,t)}\frac{\psi^{\alpha}(\rr,\vv,t)}{\na(\rr,t)}
\left( \begin{array}{ccc}
0  \\
 (\vv-\uu) \cdot  {\bf C}^{\alpha\beta}(\rr,t) \\
(\frac{\ma(\vv-\uu)^2}{3 k_B T(\rr,t)} -1)  B^{\alpha\beta}(\rr,t)  \end{array} \right) 
\ee
%%%%%%%%%%%%%%%%%%%%%
and
\be
\left( \begin{array}{ccc}
0  \\
 {\bf C}^{\alpha\beta}(\rr,t)\\
 B^{\alpha\beta}(\rr,t)\\  \end{array} \right) =
\int d\vv
\left( \begin{array}{ccc}
1   \\
\ma (\vv-\uu)   \\
\frac{\ma(\vv-\uu)^2}{2}  \end{array} \right)  \Omega^{\alpha\beta}(\rr,\vv,t).
\label{colonna}
\ee
The vanishing of the first element of the array in the l.h.s of eq. \eqref{colonna} expresses the conservation law
of the number of particles of each species in a collision.

Due to the splitting \eqref{splitting}, the orthogonal part of the collision term will not appear explicitly
in the balance equations for the hydrodynamic variables (see  eqs. \eqref{balancemass}-\eqref{balanceenergy} ).
This fact suggests  a simple approximation for  $(I-{\cal P}_{hydro})\Omega^{\alpha\beta}$, where
one replaces the exact expression by 
a BGK-like relaxation time term, having the  property
of vanishing  under the application of the
operator ${\cal P}_{hydro}$:
\be
\sum_\beta (I-{\cal P}_{hydro})\Omega^{\alpha\beta}(\rr,\vv,t)\simeq-\omega[ \fa(\rr,\vv,t)- \psi^{\alpha}_{\perp}(\rr,\vv,t)],
\label{relaxationtime}
\ee
where $\omega$ is a relaxation frequency and
\bea
&&
\psi^{\alpha}_{\perp}(\rr,\vv,t)=\psi^{\alpha}(\rr,\vv,t) \Bigl\{1+
\frac{\ma(\uua(\rr,t)-\uu(\rr,t))\cdot(\vv-\uu(\rr,t))}{k_B T(\rr,t)}\nonumber\\
&&
+\frac{\ma}{2 k_B T(\rr,t)}
\Bigl(\frac{\ma\bigl[(\uua(\rr,t)-\uu(\rr,t))\cdot(\vv-\uu(\rr,t))\bigl]^2}
{k_B T(\rr,t)}-\bigl(\uua(\rr,t)-\uu(\rr,t)\bigl)^2\Bigl)\Bigl\}. \nonumber\\
\label{prefactor}
\eea
The factor  multiplying the Maxwellian in eq. (\ref{prefactor}) 
serves as to "orthogonalize" the term $-\omega[ \fa- \psi^{\alpha}_{\perp}]$  
to the collisional terms proportional to ${\bf C}$ and $B$,
so that the BGK contribution does not explicitly affect the
balance equations.

%%%%%%%%%%%%%%%%%%%%%%%%%%

We construct, now, the hydrodynamic equations by projecting the equations \eqref{evolution}
onto the hydrodynamic subspace.
Multiplying by $\{1,\ma \vv, \ma(\vv-\uu)^2/2\}$, integrating over velocity and summing over components, we obtain the following set of balance equations:
\be
\partial_{t}\rho(\rr,t) +\nabla\cdot \Bigl(\rho(\rr,t) \uu(\rr,t) \Bigl)=0 ,
\label{balancemass}
\ee
which represents the continuity law of the mass density.
In addition we have the continuity equations for each species:
\be
\partial_{t}\rhoa(\rr,t) +\nabla\cdot \Bigl(\rhoa(\rr,t) \uua(\rr,t) \Bigl)=0 .
\label{continuity}
\ee
We also find the momentum balance equation
\bea
&&\partial_{t}[\rho(\rr,t)u_j(\rr,t)]+ \nabla_i
\left(\rho(\rr,t) u_i(\rr,t) u_j(\rr,t)]\right)\nonumber\\
&& =-\nabla_i  P^{(K)}_{ij}(\rr,t)+ \sum_{\alpha}\frac{F^{\alpha }_j(\rr)}{\ma}
\rhoa(\rr,t)+
\sum_{\alpha\beta} C^{\alpha\beta}_j(\rr,t)
\label{balancemomentum}
\eea
and the balance equation for the local temperature:
\bea
&&
\frac{3}{2}k_B n(\rr,t) \Bigl(\partial_{t} +u_i(\rr,t)\nabla_i\Bigl) T(\rr,t)
\nonumber\\
&&
=-P^{(K)}_{ij}(\rr,t) \nabla_i u_j(\rr,t)
-\nabla_i q^{(K)}_i(\rr,t)+B (\rr,t)+\sum_\alpha \rhoa(\rr,t)
\frac{F_i^\alpha(\rr)}{\ma} (u^\alpha_i(\rr,t)-u_i(\rr,t)) ,
\nonumber\\
\label{balanceenergy}
\eea
with $B=\sum_{\alpha\beta}B^{\alpha\beta}$, where
we have  introduced the kinetic part of the pressure tensor:
\be
P_{ij}^{(K)}(\rr,t)=\sum_\alpha\ma\int d\vv (\vai-u_i)(\vaj-u_j)\fa(\rr,\vv,t)
\label{pressurekin}
\ee
and the kinetic part of the heat flux vector
\be
{\bf q}^{(K)}\equiv \sum_\alpha \ma \int d\vv (\vv-\uu) \frac{(\vv-\uu)^2}{2}
\fa(\rr,\vv,t).
\label{qk}
\ee 

Eqs. \eqref{balancemomentum} and \eqref{balanceenergy} assume their
standard hydrodynamic form, if we insert the relation (see ref.\cite{Brey}) between the 
collisional moments and  the divergence of the collisional transfer contribution to the pressure:
\be
\nabla_i  P^{(C)}_{ij}(\rr,t)=-\sum_{\alpha\beta} C^{\alpha\beta}_j(\rr,t)
\label{pcollisione}
\ee
%%%%%%   CHECK
and  the relation between the collisional transfer contribution to the heat flux
and the pressure:
\be
\nabla_i q^{(C)}_i(\rr,t)+P^{(C)}_{ij}(\rr,t) \nabla_i u_j(\rr,t)=-B (\rr,t).
\label{qcollisione}
\ee

%%%%%%%%%%%%%%%%%%%%%%%%%%%%%%%%

Notice two facts:
a) at equilibrium the term \eqref{relaxationtime} vanishes,
 b) out of equilibrium  it
determines a fast relaxation
of the distributions  towards the local values of $\uu(\rr,t)$ an $T(\rr,t)$.

In order to obtain explicit expressions for ${\bf C}$ and $B$  we 
make a further approximation. We   perform  the integrals appearing in eq.\eqref{colonna}
by replacing the true distribution functions $\fa(\rr,\vv,t)$
by the  Maxwellian distributions, corresponding to average density $\na(\rr,t)$, local velocity
$\uua(\rr,t)$ and temperature $T(\rr,t)$.
Many years ago the same kind of approximation was used by Longuet-Higgins and
Pople \cite{Longuet}  to predict the transport coefficients of hard core systems.
As shown in ref. \cite{Lausanne2009}  a useful representation of the last
 term  in eq. (\ref{balancemomentum})  is obtained by the following  decomposition:
 \be
\sum_\beta{\bf C}^{\alpha\beta}(\rr,t)=\na(\rr,t)\Bigl( \FF^{\alpha,mf}(\rr,t)+\FF^{\alpha,drag}(\rr,t)+\FF^{\alpha,viscous}(\rr,t) 
+\FF^{\alpha,T}(\rr,t)\Bigl) .
\label{splitforce}
\ee
The first term in the r.h.s. can be identified with the force acting on the $\alpha$-particles at $\rr$ due to
the influence of all remaining particles in the system, the gradient  of the potential of mean force: 
\be
\FF^{\alpha,mf}(\rr,t)=-k_B T\sum_\beta\sab^2 
\int d\bk \bk
g_{\alpha\beta}(\rr,\rr+\sab\bk,t)
n_{\beta}(\rr+\sab\bk,t)+\sum_\beta \GG^{\alpha\beta}(\rr,t)
\label{mforce},
\ee
where the integral over $\bk$ is over the surface of a unit sphere.
The second term is the average drag force experienced by particles of species $\alpha$ when
moving with velocity $\ua$ with respect to the remaining species having velocities $\ub$:
\bea
\FF^{\alpha,drag}(\rr,t)&=&
-\sum_\beta 2\sab^2 \sqrt{\frac{2\muab k_B T(\rr,t))}{\pi} }\times
\nonumber\\&&
(\uua(\rr,t)-\uub(\rr,t))\cdot \int d\bk \bk \bk
g_{\alpha\beta}(\rr,\rr+\sab\bk,t)
\nb(\rr+\sab\bk,t) ,\nonumber \\
\label{dragforce}
\eea
where $\muab$ is
the reduced mass:
$
\muab=(\ma \mb)/(\ma+\mb)$.
The third term is a viscous force due to the presence of gradients in the velocity field:
\bea
\FF^{\alpha,viscous}(\rr,t)&=&
\sum_\beta 2\sab^2 \sqrt{\frac{2\muab k_B T(\rr,t))}{\pi} }\times
\nonumber\\&&
\int d\bk \bk
g_{\alpha\beta}(\rr,\rr+\sab\bk,t)
\nb(\rr+\sab\bk,t)
 \bk\cdot
(\uub(\rr+\sab\bk)-\uub(\rr)) ,\nonumber \\
\label{viscforce}
\eea
while the last term takes into account the contribution to the force due to temperature gradients:
\bea
\FF^{\alpha,T}(\rr,t)&=&
-\sum_\beta\frac{\ma}{\ma+\mb}\sab^2 
\int d\bk \bk
g_{\alpha\beta}(\rr,\rr+\sab\bk,t) \times
\nonumber\\&&
\nb(\rr+\sab\bk,t)  k_B [T(\rr+\sab\bk,t)-T(\rr,t)].
\label{thermalforce}
\eea

%%%%%%%%%%%%%%%%%%%%%%%%%%%%%%%%%%CCCCCCCCCCCCCCCCCCCCC
Finally, the energy integrals are  given by the expression:
\bea
&&B(\rr,t)= k_B T(\rr,t)\sum_{\alpha\beta} \sab^2 
\int d\bk
g_{\alpha\beta}(\rr,\rr+\sab\bk,t)\na(\rr,t) \nb(\rr+\sab\bk,t)\times\
\nonumber\\&&
\Bigl[
-\frac{\mb}{\ma+\mb}\bk\cdot[\uub(\rr+\sab\bk,t)-\uub(\rr,t)]
-\frac{\mb}{\ma+\mb}\bk\cdot[\uub(\rr,t)-\uu(\rr,t)]
\nonumber\\&&
-\frac{\ma}{\ma+\mb}\bk\cdot[\uua(\rr,t)-\uu(\rr,t)]
%\nonumber\\&&
+ \frac{2}{\ma+\mb} \sqrt{\frac{2 \muab k_B T(\rr,t)}{ \pi}}
\frac{[T(\rr+\sab\bk,t)-T(\rr,t)]}{T(\rr,t)} \Bigl]
\nonumber\\&&
+ \na(\rr,t)\GG^{\alpha\beta}(\rr,t)\cdot{\bf w}^\alpha(\rr,t),
\nonumber\\
\label{thermalb}
\eea
where ${\bf w}^\alpha(\rr,t)=\uua(\rr,t)-\uu(\rr,t)$.
Notice that eq. \eqref{splitforce} represents 
a decomposition of the total force in dissipative and non dissipative  terms.
The first ( eq. \eqref{mforce}) and the fourth term ( eq. \eqref{thermalforce})  are non dissipative and have  opposite sign
under time-reversal with respect to the partial momentum current $\na\uua$,
that they induce via the balance equation eq. (\eqref{balancemomentum}).
The dissipative terms  eq.\eqref {dragforce} and eq.\eqref{viscforce}, instead, have the same parity as the current \cite{Lubensky}.
Moreover, dissipative forces 
have an equilibrium counterpart, being  related, as shown in ref.\cite{JCP2010}, to the intrinsic chemical
potential of the individual components:
\be
\FF^{\alpha,mf}(\rr,t)= -\NN \mu_{int}^{\alpha}(\rr,t),
\label{potchimico}
\ee
while dissipative forces vanish at equilibrium.
Eq. \eqref{potchimico} represents a direct connection between the effective force featuring in the DDFT and the present
approach and this fact opens the possibility of 
transferring to systems out of equilibrium
the vast knowledge accumulated in the last twenty years concerning 
effective interactions in colloidal solutions and liquid mixtures at equilibrium \cite{biben,Roth,GED}.
On the other hand, the dissipative forces cannot be derived from a functional derivative of the
Helmholtz free energy functional, ${\cal F}(\{\na\})$.
In other words, the DDFT  does not give information about the transport
coefficients of the system, which are originated by non-equilibrium processes
not accounted for  by this approach.

Once the self-consistent fields ${\bf C}$ and $B$ have been specified, it is possible to solve eqs.
\eqref{evolution} and  study fluids under a variety of inhomogeneous conditions,
using the LBM to achieve a numerical solution of eq. \eqref{evolution}.
We have shown in a recent series of papers that the main features of a structured fluid can be captured
up to moderate packing fraction \cite{JCP2010} by a suitable extension of the LBM.
In the present paper  we shall not discuss the numerical aspects of the problem,
but investigate further some theoretical issues.

\section{Equilibrium and non equilibrium properties of bulk systems}
\label{Equilibrium properties}
In order to recover the
thermodynamic properties we impose the conditions of global equilibrium.
These require that all the  velocities $\ua$ are equal and all hydrodynamic fields 
are time independent and, with the exception of the densities
$\na$, are spatially uniform.
 With $T=$constant  $\uua=$constant, 
eq. \eqref{balancemomentum}, the momentum balance condition,  
reduces to the hydrostatic equilibrium condition.
The r.h.s. of  eq. \eqref{pcollisione} relating the equilibrium part of the 
potential contributions to the  pressure tensor
can be separated into a repulsive, $P_{ij}^{rep}\rr,t)$
and an attractive term $ P_{ij}^{attr}(\rr,t)$. The pressure can be obtained
by applying the following formulae originally derived  by Kirkwood \cite{Brey,McLennan}. 
\bea
 P_{ij}^{rep}(\rr,t)&=& \sum_{\alpha\beta}\frac{k_B T}{2}\sab^3\int d\bk \hat s_i \hat s_j 
\times\nonumber\\&&\int_0^1 d\lambda \gab(\rr-(1-\lambda)\sab\bk,\rr+\lambda\sab\bk)
\na(\rr-(1-\lambda)\sab\bk)\nb(\rr+\lambda\sab\bk)
\nonumber\\
\label{Pc}
\eea
and
\bea
 P_{ij}^{attr}(\rr,t)&=& -\sum_{\alpha\beta}\frac{1}{2}\int d^d R
\frac{ R_i R_j  }{R}
\frac{\partial U^{\alpha\beta}_{attr}(R)}{\partial R}
\times\nonumber\\&&
\int_0^1 d\lambda 
\gab(\rr+(1-\lambda) {\bf R} ,\rr-\lambda {\bf R} )\na(\rr+(1-\lambda) {\bf R} )\nb(\rr-\lambda {\bf R}) .
\nonumber
\label{Pattr}
\eea
In the case of a uniform system the total bulk pressure, $P^{bulk}$, obtained by summing the kinetic and potential contributions
and taking all the fields to be constant, is diagonal and isotropic:
\be
P^{bulk}=k_B T \sum_\alpha \na \Bigl(
1+\frac{2\pi}{3} \sum_{\beta}\sab^3\nb \gab(\sab)
- \frac{2\pi}{3k_B T}\sum_{\beta}\nb \int dR
R^3 \gab(R) \frac{\partial U^{\alpha\beta}_{attr}(R)}{\partial R}  \Bigl).
\ee
In addition, 
the surface tension of a planar interface can calculated from 
 Kirkwood and Buff formula \cite{Evans1},
 \be
 \gamma=\int_{-\infty}^{\infty} dz [P_N(z)-P_T(z)],
 \ee
 where the subscripts $N$ and $T$ indicate  
 normal and tangential  components of the pressure tensor, respectively.

{\it Non equilibrium properties.}
%\label{Non equilibrium properties}
The kinetic coefficients can be considered to be the sum of two
contributions: the first due to the instantaneous transmission of momentum and energy across the bodies 
of molecules upon collision and the second resulting from the distortion of the Maxwell-Boltzmann distribution
induced by the presence of viscous and heat flows.
Hereafter, we briefly report the calculation of the kinetic coefficients along lines similar to those
discussed in ref. \cite{Melchionna2009}.
%%%%%%%%%%%%%%%%%%%%%%%%%%% Trasporto

{\it Collisional contributions to transport coefficients.}
While kinetic transport, prevailing at low densities,  is originated by  the movement of the particles carrying a certain amount of momentum and energy, collisional transport, dominant at higher densities,
is due to  the transfer of momentum and  energy 
from one particle to the other during collisions.
Following steps similar to those of references \cite{Melchionna2008,Melchionna2009}, we derive
the collisional shear viscosity and heat conductivity of the mixture.
%%%%%%%%%%%%%%%%%%%%%%%%%%%%%%%%%%%%%
To simplify the derivation 
we assume that the densities and the temperature
are uniform ,  $n_A,n_B=const$ and $T(\rr,t)=T_0$ and $\uua=\uub=\uu$
and that the velocity field is a shear slowly varying over distances of the order of the molecular diameter:
\be
\uu(\rr,t)=(0,u_y(x,0),0).
\ee
We compute the collisional contribution to the viscosity using the relation: 
\be
\sum_{\alpha\beta}C^{\alpha\beta}_y=-\frac{\partial P_{xy}^{(C)}}{\partial x}=
\eta^{(C)} \frac{\partial^2 u_y}{\partial x^2}
\label{comp}
\ee
where we employed eq. \eqref{pcollisione} for the first equality
and the constitutive relation
\be
P^{(C)}_{xy}=P^{(C)}_{yx}=
-\eta^{(C)}[\frac{\partial u_y}{\partial x}+\frac{\partial u_x}{\partial y}]
\label{comp1}
\ee
 for the second equality.
With the help of  eqs. \eqref{splitforce} and \eqref{viscforce}, after
expanding to second order in $\sab$, we obtain:
\be
\eta^{(C)}=\frac{4}{15} \sum_{\alpha\beta} \sqrt{2\pi\muab k_B T }  \sab^4 \gab\na\nb .
\label{etac}
\ee
Similarly, if we assume $\na=constant$ , $\uua=\uub$
and small temperature gradients, we can derive the thermal conductivity. We use  the constitutive 
relation
 \be
 {\bf q^{(C)}}(\rr,t)=-\lambda^{(C)} \nabla T(\rr,t) ,
 \ee
 compare with relation \eqref{qcollisione} and
expand $B(\rr,t)$ ( formula \eqref{thermalb}) to second order in $\sab$ with the result:
\be
\lambda^{(C)}=\frac{4 }{3}k_B\sqrt{2 k_B T \pi }   \sum_{\alpha\beta} \frac{\sqrt{\muab}}{\ma+\mb} \sigma_{AB}^4  \gab\na\nb .
\label{lambdac}
\ee
%When the physical properties of the particles become identical we recover the properties of a single
%component system with density $n=n^A+n^B$ thus fulfilling the indifferentiability principle.
The present results are consistent with the theory of transport coefficients put forward for mono-disperse
hard-sphere systems
by   Longuet-Higgins and Pople more than half a century ago \cite{Longuet}.
It is based on the fact that in hard sphere systems even a local Maxwellian approximation to the distribution functions
is able to account for the collisional transfer contribution to the transport coefficients. 
According to formulae (\ref{etac}) and (\ref{lambdac})  it is clear that within the present approximation
the mean field  term,  $\GG^{\alpha\beta}$, does not contribute to the
transport coefficients.

{\it Kinetic contributions to transport coefficients.}
The self-consistent Longuet-Higgins and Pople gaussian approximation employed above
gives an expression for the transport coefficients at high densities, but does not give
an expression for the kinetic contribution to these quantities.
In order to obtain the full expression of the transport coefficients we 
apply the Chapman-Enskog analysis \cite{Chapman} .
The kinetic contribution to the transport coefficients can be derived upon neglecting
the interaction terms ${\bf C }$ and $B$ in the transport equation. 
 The solution of the transport equation  (\ref{evolution}) can be written as:
\be
\fa(\rr,\vv,t) \approx \psi^\alpha(\rr,\vv,t)+\delta\fa(\rr,\vv,t)
\label{zeroe}
\ee
In this case  eq.(\ref{evolution}) to first order in $\delta \fa$  reads:
\be
\Bigl(\partial_{t} +\vv\cdot\NN 
+\frac{\FF^{\alpha}(\rr)}{\ma}\cdot
\frac{\partial}{\partial \vv}\Bigl)     \psi^\alpha(\rr,\vv,t) =-\omega\delta\fa(\rr,\vv,t).
\label{appe2}
\ee
Substituting (\ref{zeroe}) into (\ref{appe2}) and taking the appropriate velocity moments we obtain the
following Euler-like ( i.e. without dissipative effects) hydrodynamic equations for the mixture. These are:
the continuity equation for each species
\be
\frac{\partial \na}{\partial t} + 
\frac{\partial (\na u_j)}{\partial x_j}=0,
\label{unoe}
\ee
the global momentum  conservation:
\be
\sum_{\alpha}\ma\na
[ \frac{\partial u_j}{\partial t}+ u_i  \frac{\partial u_j}{\partial x_i}]=
 -\frac{\partial P_{ij}}{\partial x_i} +\sum_\alpha F_{\alpha j}(\rr)\na(\rr,t),
\label{duee}
\ee
and  the temperature equation:
\be
\sum_{\alpha} \na(\rr,t)\Bigl( \frac{\partial T}{\partial t}+ u_i  
\frac{\partial T}{\partial x_i}\Bigl)=-
\sum_{\alpha}\frac{2}{3}\na(\rr,t) T \frac{\partial u_i}{\partial x_j}\delta_{ij}.
\label{tree}
\ee
Using again the solution (\ref{zeroe})  in eq. (\ref{appe2}) and eliminating 
the time derivatives of the hydrodynamic fields with the help of eqs. (\ref{unoe})-(\ref{tree})  
we obtain the first order correction to the distribution function:
\bea
&&
\delta \fa(\rr,\vv,t) =- \frac{1}{\omega} \psi^{\alpha} \Bigl\{\frac{(v_i-u_i)}{k_B T}\Bigl(
\frac{\ma}{\sum_\alpha \ma \na} [ 
-\frac{\partial P_{ij}}{\partial x_j} +\sum_\alpha F^{\alpha }_i
\na]
% \nonumber\\&&
+  \frac{1}{\na} \frac{\partial  (k_B T \na) }{\partial x_i}-F^{\alpha }_i \Bigl)
\nonumber\\&&
+\frac{\ma}{k_B T}\Bigl( (v_i-u_i)(v_j-u_j)  -\frac{1}{3}(\vv-\uu)^2 \delta_{ij} \Bigl)
\frac{\partial u_i }{\partial x_j} 
%\nonumber\\
+\Bigl( \ma\frac{(\vv-\uu)^2}{2}-\frac{5}{2} k_B T \Bigl) \frac{(v_i-u_i)}{k_B T^2} \frac{\partial T}{\partial x_i} \Bigl\},
\nonumber\\
 \label{quattroe}
\eea
where the first term contributes to the diffusive current,
the second term to the viscous flow and the third to the heat flow.
With the help of eq. \eqref{quattroe}, we  compute the off diagonal part of the kinetic  stress  tensor  $(i\neq j)$: 
\be
 P^{(K)}_{ij}=\sum_\alpha  m^\alpha\int d\vv\delta f^\alpha(\rr,\vv,t)[(v_i-u_i)(v_j-u_j)]
=-\eta^{(K)}  \Bigl(\frac{\partial u_i }{\partial x_j} +\frac{\partial u_j }{\partial x_i} -\frac{2}{3}\delta_{ij} \sum_l \frac{\partial u_l }{\partial x_l}
\Bigl)
 \ee
where the kinetic viscosity of the mixture is
\be
\eta^{(K)}=\frac{k_B T}{\omega}\sum_\alpha \na.
\ee
Finally, we obtain the kinetic contribution to the heat flux
\be
{\bf q}^{(K)}\equiv \sum_\alpha \ma \int d\vv (\vv-\uu) \frac{(\vv-\uu)^2}{2}
\delta \fa(\rr,\vv,t)=-\lambda^{(K)}\NN T,
\label{qk}
\ee 
where the kinetic contribution to the  heat conductivity is:
\be
\lambda^{(K)}=\frac{5 k_B^2 T }{2 \omega}\sum_\alpha \frac{\na}{\ma}.
\ee

\section{Heuristic derivation DDFT from hydrodynamics of an asymmetric binary mixture }
\label{DDFT}

Classical dynamic density functional theory  (DDFT)  is 
a widely used tool  for studying the dynamics of 
suspensions of
colloidal particles in a solvent \cite{wu,Espanol,rauscher2010,Likos,
Umbmpedromallorca}. Within this approach  only the density of the colloidal particles appears explicitly,  
 whereas the solvent appears 
through the friction coefficient, or friction tensor.
As a little remark, we wish to show that in the case of an
isothermal  hard sphere mixture in which the $c$ (for colloidal) component is very
diluted and the $s$  (for solvent) particles are light $(m^s/m^c<<1$) it is possible to derive an equation for the density
of species $c$, which is a special instance of the  DDFT equation.

Being lighter, the $s$ particles have higher thermal velocities
and  reach equilibrium faster.
It is possible to integrate out their degrees of freedom and represent
their influence by a viscous drag force, proportional to the velocity of the solute 
particles, plus a random force mimicking the effect of their random motions.

From the continuity equation \eqref{continuity} for the density of $c$  particles we have:
\be
\partial_{t}\rho^c(\rr,t)+\nabla\cdot (\uu(\rr,t)\rho^c(\rr,t))+
\nabla\cdot \frac{\rho^c(\rr,t)\rho^s(\rr,t)}{\rho(\rr,t)}(\uu^c(\rr,t)-\uu^s(\rr,t))   =0,
\label{con2}
\ee 
where we separated the advection term from the diffusion term, and $\uu^c$ and $\uu^s$ are the velocity of the
colloidal particles and of the solvent, respectively.
We now assume  that the inertial term in the momentum equation for the colloidal particles
can be neglected with respect to the molecular forces and the motion is not accelerated.
In other words, we consider a low Reynolds number regime and obtain:
 \be
-\frac{k_B T}{n^c(\rr,t)}\NN n^{c}(\rr,t)+ \FF^{c}(\rr,t) +\FF^{c,mf}(\rr,t)+\FF^{c,drag}(\rr,t)+\FF^{c,visc}(\rr,t)\simeq0
\label{bequil}
\ee
where we used eq. \eqref{splitforce}.
If, in addition, we neglect the shear force,
with the help of eqs. \eqref{dragforce}  and \eqref{bequil} we eliminate the drag force
in favour of
$(\uu^c-\uu^s)$:
\bea
\bigl(&&\uu^c(\rr,t)-\uu^s(\rr,t)\bigl) \simeq
\nonumber\\&&
-\frac{3 }{8\rho} \frac{m^s}{(2 \pi \mu_{cs} k_B T)^{1/2}\sigma_{cs}^2 
g_{cs}}
\Bigl( k_B T\NN \ln n^c(\rr,t)-\FF^{c,mf}(\rr,t) -\FF^{c}(\rr,t)\Bigl ).
\label{cdflux4}
\eea
We define
the local chemical potential, $\mu^c(\rr,t)$,  as the sum of the internal forces (see ref. \cite{JCP2010}):
$$
\NN\mu^c(\rr,t)= k_B T\NN \ln n^c(\rr,t)-\FF^{c,mf}(\rr,t),
$$
substitute into  the advection-diffusion equation (\ref{con2}),
 and take the diluted limit  $n^c/n^s<<1$ and $\uu\simeq \uu^s$:
\bea
%&&
\partial_{t}n^c(\rr,t)+\nabla\cdot (\uu(\rr,t) n^c(\rr,t))=
%\nonumber\\&&
\frac{1  }{\gamma}\nabla\cdot \Bigl[ n^c(\rr,t)
\Bigl( \NN \mu^{c}(\rr,t) -\FF^{c}(\rr)\Bigl)\Bigl],
\nonumber\\
\label{ddft}
\eea 
where the friction coefficient,  $\gamma$,  is given by the expression:
\be
\frac{1}{\gamma}=\frac{3}{8 n^s}\frac{1}{ \sqrt {\pi m^s k_B T} \sigma_{cs}^2 g_{cs} } .
\label{dab}
\ee
We, now, observe that eq.\eqref{ddft} 
is formally identical to a  DDFT equation for the $c$ species in a velocity field $\uu(\rr,t)$.
As $n^c\to 0$ the gradient of the chemical potential $\mu^c$  approaches the ideal gas value, $k_B T \NN n^c(\rr,t)$,
so that  eq. \eqref{ddft} becomes a linear advection-diffusion equation for the field $n^c$, with a diffusion coefficient
given by:
\be
D= \frac{k_B T}{\gamma},
\ee
to be interpreted as a fluctuation-dissipation relation between $\gamma$ and $D$ .
 Apparently, such a result sounds correct, appealing and satisfactory, but
contradicts Einstein's  theory of Brownian motion stating that the diffusion coefficient  for a large and massive 
solute molecule of diameter $\sigma_{cc}$ immersed in a solvent of much smaller and 
lighter molecules is related to the solvent viscosity by the 
Stokes-Einstein relation:
\be
D_{SE}= \frac{k_B T}{\gamma_{hydro}}.
\ee
Einstein's theory relies on the hypothesis that a diffusing spherical body moves among solute particles
as a macroscopic sphere does in a viscous
incompressible continuum fluid. 
One first needs to solve the Stokes
equation for the flow of the fluid around the sphere, with the condition of  no
slip at its surface. From the flow one computes  the total stress acting on the sphere
surface and finally the drag force, which results proportional
to the (hydrodynamic)  radius, $\sigma_{cc}/2$, of the object and to the viscosity, $\eta_s$, of the solvent
according to the Stokes formula:
 \be
\FF^{c,drag}=-\gamma_{hydro}(\uu^c-\uu)=-3\pi\eta_s\sigma_{cc}(\uu^c-\uu).
\label{ghydro}
\ee

The Enskog  relation \eqref{dab} and the Stokes-Einstein relation \eqref{ghydro}  display different scalings with respect to the 
physical properties  of the $c$ particles.
The relation
$$
\gamma_{hydro}=3\pi\eta_s\sigma_{cc}
$$
 for the friction coefficient shows
 a linear dependence
on the solute diameter $\sigma_{cc}$, but no dependence on its mass,
while the Enskog approximation considerably underestimates the friction coefficient, which
depends on the reduced mass of the solute-solvent pair and
 has a quadratic dependence on the solute diameter.
Only for very heavy solutes the mass dependence of Enskog friction coefficient
vanishes (see eq. \eqref{cdflux4}). The numerically computed friction coefficient shows a crossover to the Stokes-Einstein
result only for large values of the mass ratio at fixed size ratio.
A full merging between the two approaches, in spite of repeated attempts, has not been achieved so far \cite{Kapral,Schmidt,Sung}.

Clearly, the Enskog microscopic picture, where the interactions occur via collisions with the solvent
particles, and the Einstein picture, which treats  the solvent molecules  as a continuum and the
 effect of the solute-solvent interactions by means of
the no slip boundary condition, account for different mechanisms. 
Einstein's theory is correct for large  
solute particles, where the
typical ratios between the radii of suspended colloidal particles and solvent molecules
range in the interval $30-3000$ and the mass ratio is in the interval $10^{4}-10^{10}$.  
These large differences justify the the idea of complete separation of time scale between the motion of the
solute, which occurs in a time of the order of $\tau_c=m^c/(3\pi\eta^s\sigma_{cc})$, and the motion of the solvent
characterized by the Enskog collision time
$\tau_s\simeq n^s \sigma_{ss}^2 \sqrt{k_B T/m^s}$, the first being 5 orders of magnitude
larger \cite{hansenbocquet}.
However, for small colloids the collisional and hydrodynamic
regimes are not well separated.  A crossover should occur when
the size and the mass of the solute become comparable to those of the solvent molecules.
In this case eq. \eqref{ghydro} is not expected to be valid. 
The short time direct collisions between the tracer particle and the solvent particles 
are indeed  taken into account  by Enskog's theory, which, 
however, misses the long time behavior  determined 
by correlated re-collisions of the tracer
with the same solvent molecule \cite{Musharaf}.
The velocity auto-correlation function (VACF) of the $c$ particles in Enskog's theory
displays an exponential decay which is valid in the short time region, but does not show
the algebraic inverse power law tails  characterizing the long time behavior of the VACF.
The pure  exponential behavior is the result of the  molecular-chaos assumption, i.e. of a Markovian representation of the dynamics,
well describing the large mean free path regime. The RET is able to capture  the formation of positional correlations, but not
the build-up of velocity correlations among the particles, at the origin of the Stokes-Einstein behavior of the diffusion coefficient.
It is our opinion that methods, which truncate the BBGKY hierarchy at the level of one-particle distribution functions,
are not apt to account for the renormalization of the transport coefficients due to the
presence of velocity correlations, the ultimate cause of hydrodynamic interactions.
With respect to this problem mode coupling theory (MCT) is able
to predict the correct behavior by including at the same time binary collision processes, coupling 
with  transverse  current of the solvent and with  density fluctuations \cite{Bagchi}.

Finally, it is worth to mention that a simpler way to bridge 
Einstein with Enskog theory as the size or the colloids is reduced to that of the solvent particles
has been described in ref. \cite{hansen,berne} , where the motion of the colloid 
is described by a Langevin equation  
with a friction coefficient which is non-local in time in order to reproduce a non Markovian behavior.
Although, such an approach is frankly phenomenological it is able to predict the backscattering effect.

%The application to inhomogeneous systems of MCT perhaps in conjunction with elements of DDFT
%represents a fascinating challenge.

\section{Conclusions}
\label{Conclusions}

In this paper we have considered how microscopic methods, utilized to study  
the non-equilibrium evolution of colloidal solutions and well
accounting for  their configurational properties, 
 can be extended  to investigate the dynamics of liquid mixtures. 
We have adopted a strategy based upon kinetic methods
and shown that the resulting equations  correctly describe the hydrodynamic behavior of the density, momentum and energy transport.
We have found similarities with the DDFT equations, and this not surprising since 
the common basic ingredient of both methods is the free energy  which in one way or another 
features in the theory.
In DDFT we directly use the free energy functional, and in the kinetic approach we employ
its functional derivative,
the inhomogeneous positional pair correlation function, $g(r,r')$.
Finally, to link further the two descriptions we have introduced an
heuristic derivation of the DDFT equation from the kinetic equations for a hard-sphere mixture. Our result
seems to indicate the possibility of extending this procedure to mixture  of  higher physical interest.   

\section*{Acknowledgments}

It is a pleasure to dedicate this paper to Bob Evans, whose work has deeply influenced the development of
condensed matter in the last forty years. I had the privilege to work with him in the H.H. Wills Physics Laboratory  at the University of  Bristol.

\end{document}